\documentclass[preprint,showpacs,preprintnumbers,amsmath,amssymb]{revtex4}

\usepackage{graphicx}
\usepackage{dcolumn}
\usepackage{bm}

\begin{document}

\preprint{}

\title{Simultaneous slow phase and group velocities of light in an anisotropic metamaterial}
\author{Hailu Luo} \thanks{Author to whom correspondence should be addressed.
E-mail: hailuluo@sohu.com}
\author{Weixing Shu}
\author{Fei Li}
\author{Zhongzhou Ren}
\affiliation{Department of Physics, Nanjing University, Nanjing
210008, China}
\date{\today}

\begin{abstract}
We study theoretically the effect of ultraslow phase and group
velocities in an anisotropic metamaterial. The ultraslow phase
propagation is induced by the hyperbolic dispersion relation.
While the inherent physics underlying the slow group velocity are
collective operations of the frequency and spatial dispersion. We
show that a Gaussian wave packet exhibits simultaneous slow phase
and group velocities which depend on the choice of incident angles
and principal axis angles. The anisotrocpic metamaterial slab can
be constructed and the ultraslow phase and group velocities can be
measured experimentally.
\end{abstract}

\maketitle

\section{Introduction}\label{Introduction}

The unusual properties of the electromagnetic waves in a medium
with simultaneously permittivity $\varepsilon$ and the
permeability $\mu$ were firstly introduced by Veselago about forty
years ago~\cite{Veselago1968}. The waves propagation in such media
are quite counterintuitive. For example, the direction of energy
flow for a plane wave is opposite to the direction of propagation.
After the first experimental observation using a metamaterial
composed of split ring resonators
(SRR)~\cite{Smith2000,Shelby2001}, the study of such materials has
received increasing attention over the last few years. As noted
earlier, both $\varepsilon$ and $\mu$ are necessarily frequency
dispersive in LHM. Since the frequency dispersion is important,
the superluminal~\cite{Ziolkowski2001,Woodley2004,Gupta2004} and
the subluminal
propagation~\cite{Gupta2004,Gennaro2005,Dolling2006} in the LHM
takes place.

It should be noted that we neglect bianisotropic effects despite
the fact that they might be important for the characterization of
some rings. While negative refraction is most easily visualized in
an isotropic metamaterial, negative refraction can also be
realized in anisotropic metamaterial, which does not necessarily
require that all tensor elements of $\boldsymbol{\varepsilon}$ and
$\boldsymbol{\mu}$ have negative
values~\cite{Lindell2001,Parazzoli2003,Smith2003,Smith2004,Thomas2005,Luo2005}.
If we just consider the spatial dispersion, the superluminal group
velocity can be expected in the AMM with hyperboloid dispersion
relation~\cite{Luo2006}. While the real AMM constructed by SRR is
highly dispersive, both in spatial and frequency
sense~\cite{Parazzoli2003,Smith2003,Smith2004,Thomas2005}, hence
It is very necessary to extend the previous work and take the
frequency dispersion into account.

In the present letter, we will investigate the simultaneous slow
phase and group velocities of wave packet in an anisotropic
metamaterial. The subluminal phase propagation is induced by the
hyperbolic dispersion relations associated with the AMM.  We
describe a modulated Gaussian wave packet incident on the
anisotropic metamaterial, which demonstrates in a straightforward
manner that the peak of the localized wave packet displays
interesting smultaneous slow phase and group velocities.

\section{Hyperbolic dispersion relation}\label{sec2}
It is currently well accepted that a better model is to consider
anisotropic constitutive parameters, which can be diagonalized in
the coordinate system collinear with the principal axes of the
metamaterial. If we take the principal axis as the $z$ axis, the
permittivity and permeability tensors have the following forms:
\begin{eqnarray}
\boldsymbol{\varepsilon}=\left[
\begin{array}{ccc}
\varepsilon_x (\omega)  &0 &0 \\
0 & \varepsilon_y (\omega)  &0\\
0 &0 & \varepsilon_z (\omega)
\end{array}
\right],
\end{eqnarray}
\begin{eqnarray}
\boldsymbol{\mu}=\left[
\begin{array}{ccc}
\mu_x (\omega)  &0 &0 \\
0 & \mu_y(\omega) &0\\
0 &0 & \mu_z (\omega)
\end{array}
\right],\label{matrix}
\end{eqnarray}
where $\varepsilon_i$ and $\mu_i$  are the relative permittivity
and permeability constants in the principal coordinate system
($i=x,y,z$). It should be noted that the real AMM constructed by
SRR is highly dispersive, both in spatial sense and frequency
sense~\cite{Parazzoli2003,Smith2003,Smith2004,Thomas2005}. So
these relative values are functions of the angle frequency
$\omega$. A general study on the shape of the dispersion relation
as function of the sign of these parameters has already been
offered in Ref.~\cite{Smith2003}. In this work, we are interested
in the case of AMM with hyperbolic dispersion relation.

\begin{figure}
\includegraphics[width=10cm]{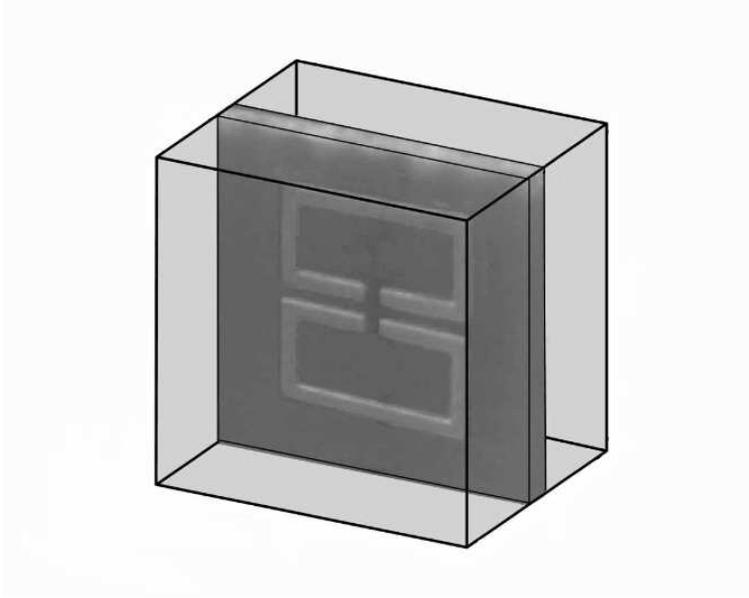}
\caption{\label{Fig1} The unit cell of the AMM is fabricated by
printing rows of the SRRs onto large sheets of the dielectric
substrate. The remaining volume is air-filled. For the sake of
numerical calculations we choose the same parameters in the
experiment carried out in Ref.~\cite{Thomas2005} }
\end{figure}

Without loss of generality, we assume that the wave vector locate
at the $x-z$  plane ($k_y=q_y=0$).  Maxwell's equations yield a
scalar wave equation for E-polarized field . In free space, the
accompanying dispersion relation has the familiar form:
\begin{equation}
 k_{x}^2+k_{z}^2= \frac{\omega^2}{c^2}, \label{D1}
\end{equation}
where $k_x$ and $k_z$ are the $x$ and $z$ components of the
incident wave vector, $\omega$ is the frequency, and $c$ is the
speed of light in vacuum. We assume that there is an angle
$\varphi$ between the principal axis and the $z$ axis. For the
given polarization, the waves equation yield the dispersion
relation in AMM as
\begin{equation}
\alpha q_{x}^2+\beta q_{z}^2+\gamma
q_{x}q_{z}=\frac{\omega^2}{c^2}, \label{D2}
\end{equation}
where $q_{x}$ and $q_{z}$ represent the $x$ and $z$ components of
refracted wave vector, $\alpha$, $\beta$ and $\gamma$ are given by
\begin{eqnarray}
\alpha &=&\frac{1}{ \varepsilon_x \varepsilon_z \mu_y }
(\varepsilon_x \cos^2\varphi+\varepsilon_z \sin^2\varphi),\nonumber\\
\beta &=&\frac{1}{ \varepsilon_x \varepsilon_z \mu_y }
(\varepsilon_x \sin^2\varphi+\varepsilon_z \cos^2\varphi),\nonumber\\
\gamma &=&\frac{1}{\varepsilon_x \varepsilon_z \mu_y}
(\varepsilon_z \sin 2\varphi -\varepsilon_x \sin 2\varphi).
\end{eqnarray}
The material axes are normal to the surface of the rings. The cell
structure is shown in Fig.~\ref{Fig1}. The metamaterial is formed
by cutting the sheet into strips which are lined up with the
appropriate periodicity. The permeability, $\mu_z$, can be
approximated by the Lorentz model
\begin{equation}
\mu_z =
1-\frac{\omega_{mp}^2-\omega_{mo}^2}{\omega^2-\omega_{mo}^2-i
\Gamma_m \omega }, \label{uz}
\end{equation}
where $\omega_{mo}$, $\omega_{mp}$, and $\Gamma_m$ denote the
magnetic resonate frequency, plasma frequency, and damping
frequency, respectively. The characteristic resonance frequency
$f_{mo}=10.08 $GHz, plasma frequency $f_{mp}=10.56 $GHz and
$\Gamma_m=0.1$GHz ($f=\omega/2\pi$). Ignoring the metallic
structure, $\varepsilon_y$ and $\mu_x$ assume the values of the
background material which is dominantly air leading us to use
$\varepsilon_y=\mu_x=1$~\cite{Thomas2005}.

\begin{figure}
\includegraphics[width=10cm]{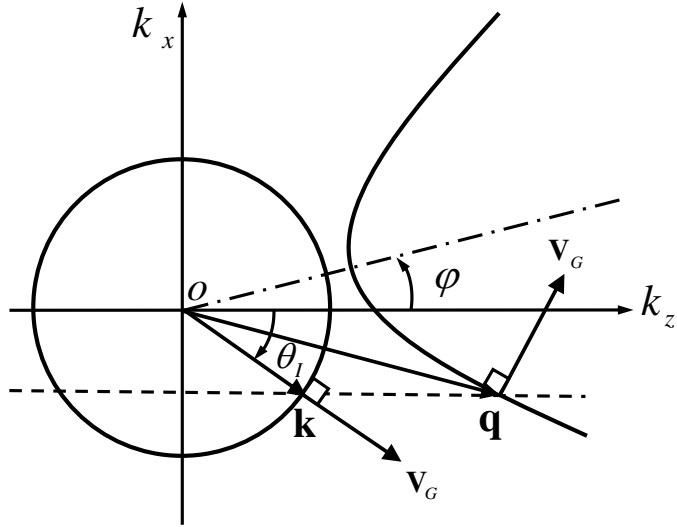}
\caption{\label{Fig2} The circle and the hyperbola represent the
dispersion relations of free space and AMM, respectively. The
frequency contour is rotated by an angle $\varphi$. The incident
wave vector ${\bf k}$ is parallel to the group velocity ${\bf
v}_G$ in free space. Because of the anisotropy, the group velocity
${\bf v}_G$ in the AMM is not necessarily parallel to the
refracted wave vector ${\bf q}$. }
\end{figure}

We assume here that the electric field is polarized along the $y$
axis. We choose to  $\mu_z(\omega)<0$ we find the corresponding
requency contour is a hyperbola as shown in Fig.~\ref{Fig2}, where
a plane electromagnetic wave is incident from free space into the
AMM. We choose the $z$ axis to be normal to the interface, the $x$
axis in the plane of the figure, and the $y$ axis out of the plane
of the figure. Due to the angular dispersion, the transmitted wave
components may refract at slightly different angles. The values of
refracted wave vector can be found by using the boundary condition
and hyperbolic dispersion relation~\cite{Luo2005}.  The
$z$-component of the wave vector can be found by the solution of
Eq.~(\ref{D2}), which yields
\begin{equation}
 q_z = \frac{1}{2 \beta}\bigg[\sqrt {4\beta \frac{\omega^2}{c^2}+(\gamma^2-4
\alpha \beta )q_x^2}-\gamma q_x\bigg], \label{qz}
\end{equation}
the choice of sign of $q_z$ ensures that light power propagates
away from the surface to the $+z$ direction.

We now determine the angle of phase refraction. The incident angle
of light in free space is $\theta_I =\tan^{-1}[k_{x}/k_{z}]$ and
the refraction angle of the transmitted wave vector in AMM can be
found by $\theta_T= \tan^{-1}[q_{x}/q_{z}]$. From the boundary
condition at the interface $z=0$, the tangential components of the
wave vectors must be continuous, i.e., $q_{x}=k_{x}$. Thus the
refracted angle of wave vector in the AMM can be easily obtained.
When the interface is aligned an angle with the optical axes of
the AMM, the hyperbolic dispersion relations will exhibit some
interesting effects.

\begin{figure}
\includegraphics[width=10cm]{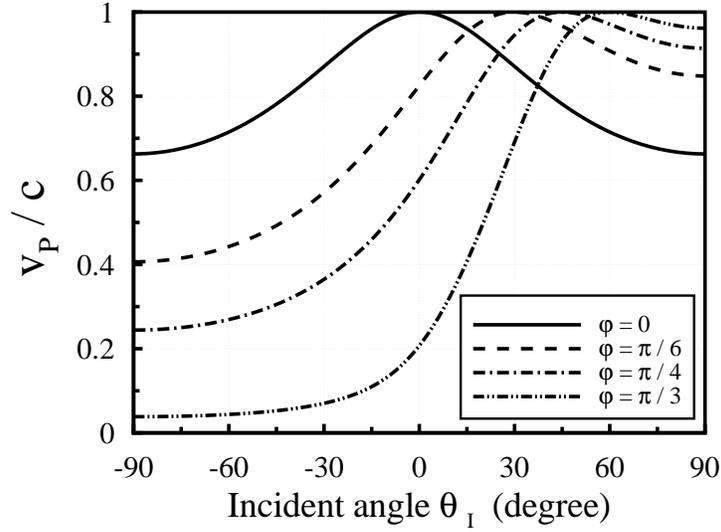}
\caption{\label{Fig3} The slow phase velocity in an AMM with
different principal axial angles of $\varphi=0$ (solid), $\pi/6$
(dashed), $\pi/4$ (dash-doted) and $\pi/3$ (dash-dot-doted). The
frequency of the incident wave is $f=10.15$GHz.}
\end{figure}

To get a deeper insight into the slow phase velocity, we plot the
frequency contour in Fig.~\ref{Fig2}. The magnitude of ${\bf q}$
varies as a function of its direction. When the refractive wave
vector ${\bf q}$ is approximately parallel to the asymptotic line
of hyperbola, ${\bf q}$ can be very large. The wave front travels
in the AMM with the velocity of $v_p=\omega/q$, so the ultra-slow
phase velocity can be expected. The slow phase velocity in an AMM
with different principal axial angles as shown in Fig.~\ref{Fig3}.

It should be noted that a wave group can be  formed  from plane
waves with different frequencies or form plane waves with
different wave vectors. Thus the difference in the phase and group
velocities can be caused by media that are dispersive or
anisotropic~\cite{Kong1990}. The group velocity is a very
important physical quantity because it identifies the speed of the
maximum intensity of wave packet. The group velocity in
anisotropic media can be defined as
\begin{equation}
{\bf v}_{G}=-Re\left[\frac{\partial D /\partial q_x}{ \partial D
/\partial \omega}\right]{\bf e}_x-Re\left[ \frac{\partial D
/\partial q_z}{
\partial D /\partial \omega}\right]{\bf e}_z, \label{vg}
\end{equation}
where $D(q_x,q_z,\omega)=\alpha q_{x}^2+\beta q_{z}^2+\gamma
q_{x}q_{z}-\omega^2/c^2$, ${\bf e}_x$ and ${\bf e}_z$ are unit
cartesian vectors. Because of the anisotropy and angular
dispersion, the group velocity is not necessarily parallel to the
wave vector ${\bf q}$. Due to the resonance effect below the
magnetic plasma frequency, the permeability functions undergo
large changes with frequency, which results in $\partial D/
\partial \omega \gg  \partial D/ \partial q_x$ and $\partial D/
\partial \omega \gg  \partial D/ \partial q_z$, hence the
subluminal group velocity can be deduced.

\begin{figure}
\includegraphics[width=10cm]{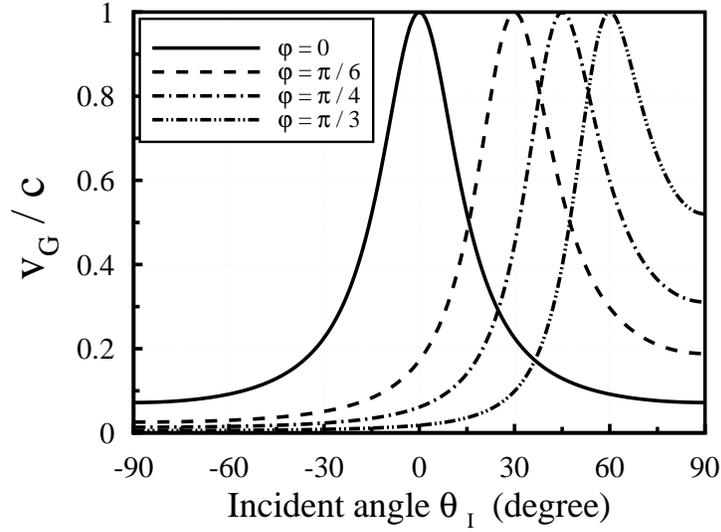}
\caption{\label{Fig4} The subluminal group velocity in an AMM with
different optical principal angles of $\varphi=0$ (solid), $\pi/6$
(dashed), $\pi/4$ (dash-doted) and $\pi/3$ (dash-dot-doted). The
frequency of the incident wave is $f=10.15$GHz.}
\end{figure}

In Fig.~\ref{Fig4}, the group velocities in the AMM with different
principal axes angle are plotted. It should be noted that the slow
group velocity in the AMM is induced by the hyperbolic dispersion
relation. While the inherent physics underlying the slow group
velocity are collective operations of the frequency and spatial
dispersion.

\section{Subluminal group propagation}\label{sec3}
To obtain a better physical picture of subluminal group velocity
in AMM, a modulated Gaussian wave packet of finite width can be
constructed. The field intensity distribution in free space is
obtained by the Fourier integral and angular spectrum
representation. Following the method outlined by Lu {\it et
al.}~\cite{Lu2004}, let us consider a modulated Gaussian
wavepacket is incident from free space
\begin{equation}
E_I(x, z) = \int_{-\infty}^{+\infty}d^2 k_\perp f( {\bf k}-{\bf
k}_0) \exp[i({\bf k} \cdot {\bf r}- i \omega({\bf k}) t],
\label{EI}
\end{equation}
where $\omega({\bf k})=c k$. we assume its Gaussian weight is
\begin{equation}
f (k_x, k_z) = \frac{a^2}{\pi} \exp [- a^2 (k_x^2+k_z^2)],
\label{ft}
\end{equation}
where $a$ is the spatial extent of the incident wave packet. We
want the Gaussian wave packet to be aligned with the incident
direction defined by the vector ${\bf k}_0=k_0 \cos \theta_I {\bf
e}_x+ k_0 \sin \theta_I {\bf e}_z$, which makes an angle
$\theta_I$ with the surface normal.

Due to the angular dispersion and the anisotropy, the transmitted
wave components may refract at slightly different angles. When the
Gaussian wave packet enters the AMM, it will no longer maintain
Gaussian, but becomes a tilted wave packet. Matching the boundary
conditions for each $k$ component at $z=0$ gives the complex field
in the form
\begin{equation}
E_T(x, z) = \int_{-\infty}^{+\infty}d^2 k_\perp f( {\bf k}-{\bf
k}_0)T ({\bf k}) \exp[i({\bf q} \cdot {\bf r}- i \omega({\bf q})
t], \label{ET}
\end{equation}
where $\omega({\bf q})=c q$ and $T ({\bf k})$ is the transmission
coefficient.

\begin{figure}
\includegraphics[width=15cm]{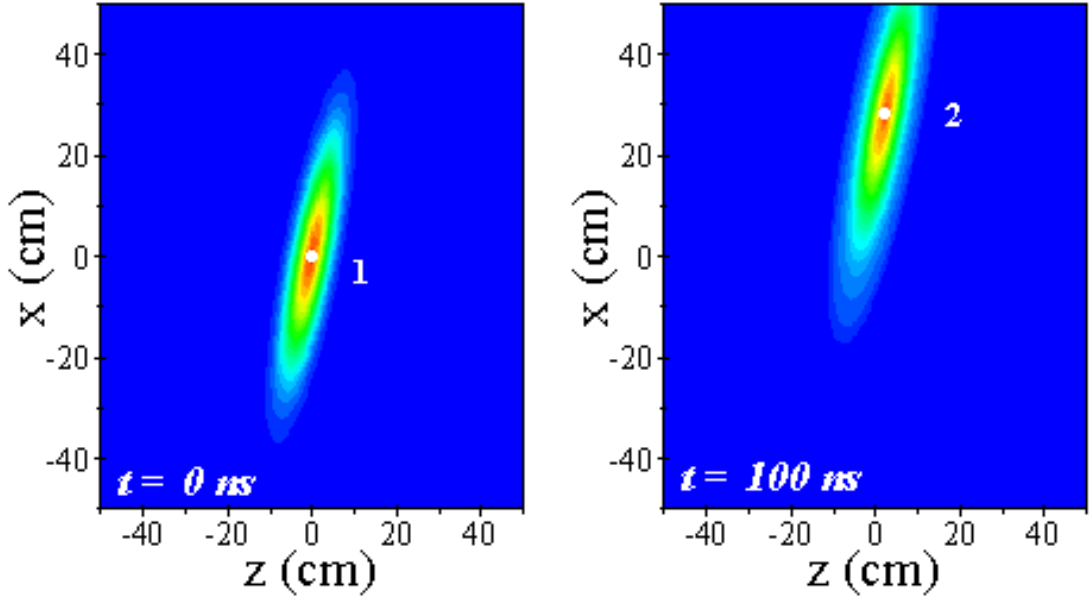}
\caption{\label{Fig5} The temporal evolution of the Gaussian wave
packet in the AMM. The center frequency is $f=10.05 $GHz with the
incident angle $\theta_I=-\pi/6$. The optical axis angle
$\varphi=\pi/3$ and spatial extent of the incident wave packet is
$a=10cm$. The peak of localized wave packet travels from point $1$
to point $2$ during the propagation time  $\Delta t$ with group
velocity $v_G \approx c/100$.}
\end{figure}

A close look at the tilted wave packet shows that the subluminal
propagation of the peak is induced by the hyperbolic dispersion
relation. Fig.~\ref{Fig5} shows a closes view of the field
intensity distribution of the wave packet propagating in the AMM.
In a propagation time $\Delta t$, the peak of the tilted Gaussian
packet travels from point $1$ to point $2$. In Fig.~\ref{Fig5} we
set the center wave vector with a incident angles
$\theta_I=-\pi/6$. We mark on each position of the peak of wave
packet at each of the two times. In the $\Delta t=100 ns$, the
peak of wave packet moves from $(0,~0)$ to $(2.45,~28.64)$. This
propagating velocity corresponds to $28.75 cm/ns$, which is almost
exactly the analytical group velocity $v_G \approx c/100$ in
Eq.~(\ref{vg}), and the slow group velocity is demonstrated
theoretically.

\section{Discussion and conclusion}\label{sec4}
In summary, we have discussed simultaneous slow phase and group
velocities of light in an anisotropic Metamaterial. It should be
noted that the slow group velocity has completely different origin
from those described in isotropic LHM~\cite{Gupta2004,Gennaro2005}
or ultracold gas of atoms~\cite{Harris1999,Hau1999}, where the
slow group velocity is caused by the frequency dispersion of the
medium. In the case discussed here, the inherent physics
underlying the slow group velocity are collective operations of
the frequency and spatial dispersion. As far as we know, this kind
of slow group velocity has not been recognized before. It should
be mentioned, however, that the shape of Gaussian wave packet is
distorted once it is incident into the AMM. It is shown that the
simultaneously slow phase and group velocities can be measured
experimentally.

\begin{acknowledgements}
This work was supported by projects of the National Natural
Science Foundation of China (No. 10125521 and No. 10535010), and
the 973 National Major State Basic Research and Development of
China (No. G2000077400).
\end{acknowledgements}

\end{document}